\documentclass[runningheads]{llncs}
\usepackage[T1]{fontenc}
\usepackage{graphicx}
\usepackage{subfiles}
\usepackage{xcolor}
\usepackage{amssymb}
\usepackage{amsmath}
\usepackage{algorithm}
\usepackage{algpseudocode}
\usepackage{listings}%
\usepackage[noend]{algcompatible}
\usepackage{hyperref}
\usepackage{rotating}
\usepackage{tabularray}
\usepackage{caption}
\usepackage{subcaption}
\usepackage{array, tabularx, booktabs}
\usepackage{float}
\usepackage{placeins}
\usepackage{multirow}

\begin{document}
\title{Semantic Labeling for Third-Party Cybersecurity Risk Assessment: A Semi-Supervised Approach to Intent-Aware Question Retrieval}

\titlerunning{Semantic Labeling for TPRA Question Retrieval}

\author{
Ali {Nour Eldin}\inst{1}\orcidID{0000-0002-0956-8108}\thanks{Corresponding author}
\and Mohamed Sellami\inst{1}\orcidID{0000-0002-7547-1857}
\and Mehdi Acheli\inst{1}\orcidID{0000-0001-9649-7127}
\and Walid Gaaloul\inst{1}\orcidID{0000-0003-0451-532X}
\and Julien Steunou\inst{2}
}

\authorrunning{A. {Nour Eldin} et al.}

\institute{
SAMOVAR, Telecom SudParis, Institut Polytechnique de Paris, Palaiseau, France\\
\email{firstname.last\_name@telecom-sudparis.eu}
\and
Board of Cyber, France\\
\email{jsteunou@boardofcyber.io}
}

\maketitle

\begin{abstract}
Third-Party Risk Assessment (TPRA) relies on large repositories of cybersecurity compliance questions used to assess external suppliers against standards such as ISO/IEC 27001 and NIST. In practice, not all questions are relevant for a specific supplier and selecting questions for a given assessment context remains a manual and time-consuming task. Existing question retrieval approaches based on lexical or semantic similarity can identify topically related questions, but they often fail to capture the underlying assessment intent, including control domain and evaluation scope.

To address this limitation, we investigate whether an explicit semantic label space can improve intent-aware TPRA question selection. In particular, we separate label space discovery from large-scale label assignment. We start by discovering overlapping clusters of semantically similar questions and then exploit LLMs to assign unique labels for each cluster. Second, we propagate labels through k-nearest neighbors (kNN) for a larger-scale question annotation. Question retrieval is finally achieved by similarity measure of the query with respect to the extracted labels instead of the questions themselves.
This reduces repeated LLM calls while preserving label consistency.

Experimental results show that the proposed semi-supervised framework reduces labeling cost and runtime compared with per-question LLM annotation while maintaining label quality and improving efficiency. Furthermore, label-based retrieval achieves better alignment with cybersecurity control domains and assessment scope than similarity-based retrieval, highlighting the value of semantic labels as an intermediate representation.

\keywords{

Third-Party Risk Assessment \and Cybersecurity Compliance \and Semantic Labeling \and Information Retrieval \and Semi-Supervised Learning \and Large Language Models \and Question Retrieval}
\end{abstract}

\section{Introduction}
Modern Security Operations Centers (SOCs) operate in an increasingly complex cybersecurity landscape and rely on diverse sources of security knowledge to support monitoring, analysis, and response. These sources include operational data such as logs and alerts, as well as more structured knowledge such as audit artifacts, compliance documentation, and governance information~\cite{DBLP:conf/acsac/HusariAACN17}. Efficiently organizing and retrieving this knowledge is essential for timely and informed security decisions.

One procedure for collecting such knowledge is Third-Party Risk Assessment (TPRA). It plays an important role in helping organizations assess the security posture of external providers such as cloud platforms and Software-as-a-Service (SaaS) vendors~\cite{keskin2021cyber,osunji2021know}. TPRA supports vendor risk profiling, supply chain security, and broader risk-aware decision-making by providing structured evidence about external security dependencies~\cite{boardofcyber_tprm}. From this perspective, TPRA can be seen as a proactive complement to SOC activities, providing assessment knowledge that complements later monitoring and response activities.

In practice, TPRA is commonly performed using structured questionnaires derived from widely adopted standards and frameworks such as ISO/IEC 27001 and NIST~\cite{iso27001,nist80053}. Organizations typically maintain internal repositories of assessment questions and must select the most relevant subset according to a supplier’s service scope, risk profile, and business context~\cite{keskin2021cyber,osunji2021know,boardofcyber_tprm}. This step is important because it directly affects the quality of the assessment and influences downstream decisions such as onboarding, remediation, and risk treatment. For example, a first supplier, an SaaS HR platform, will be questioned on data privacy, cloud security and access control. Conversely, a second one, a hardware maintenance provider, will be assessed concerning physical access, operational continuity and insider risk. Practically however,  as these repositories grow over time, question selection becomes increasingly manual, repetitive, and time-consuming. In addition, repositories often accumulate redundant questions, inconsistent phrasing, and partial semantic overlap, which makes reuse more difficult and increases the effort required from domain experts.

A natural way to address this problem is to formulate question selection as an information retrieval task. Existing approaches use lexical or semantic similarity to rank candidate questions. Traditional methods such as BM25 rely on vocabulary overlap and are sensitive to phrasing variation, while embedding-based approaches support broader semantic matching beyond exact wording~\cite{robertson2009bm25,wang2020measurement,DBLP:conf/www/SteckEK24}. Yet these methods primarily operate at the level of textual similarity and do not explicitly represent the underlying assessment intent of a question. In the TPRA setting, relevance depends not only on topical relatedness, but also on whether questions align with the same control domain and evaluation scope. For example, the questions ``Q1: Is there a documented access control policy?'' and ``Q2: Are user access rights reviewed periodically?'' both belong to the access control domain, but they differ in evaluation scope: the former asks about policy definition, while the latter concerns control review in practice. As a result, similarity-based retrieval may return questions that are semantically related but operationally misaligned with the intended assessment objective.

This limitation highlights the need for an intermediate representation that more explicitly captures TPRA question meaning than raw text similarity. In particular, such a representation should encode shared control intent in a way that allows for scalable reuse across large repositories while remaining understandable to security professionals. Shared semantic labels are a promising candidate for this purpose because they can clarify assessment intent and support retrieval at the meaning level rather than just phrasing. This challenge relates to retrieval-augmented systems, where the quality and structure of retrieved knowledge directly affect downstream reasoning~\cite{DBLP:conf/nips/LewisPPPKGKLYR020}.

To address this need, we propose a hybrid Semi-Supervised Semantic Labeling (SSSL) framework for structuring and retrieving TPRA questions. The framework combines embedding-based clustering with selective Large Language Model (LLM)-assisted annotation to induce an interpretable semantic label space, and applies k-nearest neighbors (kNN)-based label propagation to extend labeling efficiently across larger repositories. This design separates label discovery from large-scale label assignment, reducing repeated LLM inference while preserving scalability and label consistency. The resulting semantic label space is used to support more intent-aware retrieval of assessment questions.

Accordingly, this paper addresses the following research questions:
\begin{itemize}
    \item \textit{RQ1}: Does introducing an explicit semantic label space improve the alignment of retrieved TPRA questions with control domains and assessment scope compared to similarity-based retrieval?
    \item \textit{RQ2}: Can a hybrid semi-supervised labeling approach reduce reliance on Large Language Models (LLMs) while maintaining label quality and generalization performance?
    \item \textit{RQ3}: How does the proposed approach balance labeling accuracy, computational cost, and scalability compared to LLM-only methods?
\end{itemize}

The main contributions of this paper are threefold. First, we introduce a hybrid semi-supervised semantic labeling framework for structuring TPRA repositories through embedding-based clustering and selective LLM-assisted annotation. Second, we show how the induced semantic labels can be used as an intermediate representation for more intent-aware retrieval of cybersecurity assessment questions. Third, we analyze the trade-off between retrieval alignment, label quality, and computational efficiency relative to LLM-only labeling approaches.

The remainder of the paper is organized as follows. Section~\ref{sec:related_work} reviews related work on TPRA automation, retrieval methods, and semantic labeling. Section~\ref{sec:approach} presents the proposed SSSL framework. Section~\ref{sec:evaluation} describes the experimental setup and results. Section~\ref{sec:conclusion} concludes the paper and outlines future directions.

\section{Related Work}
\label{sec:related_work}

Recent work on TPRA automation spans both industrial systems and academic research. We group related work into three areas: TPRA systems, similarity-based retrieval methods, and text labeling approaches.

\subsection{Third-Party Risk Assessment Systems}
TPRA is commonly implemented within third-party risk management (TPRM) workflows that standardize vendor onboarding, questionnaire-based assessments, remediation tracking, and continuous monitoring~\cite{keskin2021cyber}. Commercial platforms, such as Board of Cyber~\cite{boardofcyber_tprm}, OneTrust~\cite{onetrust_tprm}, and others, support these workflows through curated question libraries, predefined mappings to compliance frameworks, and reporting pipelines. In this broader context, TPRA contributes to SOC activities by providing structured insights into external dependencies that inform risk-aware monitoring and decision-making.

While effective for governance and auditability, these systems rely heavily on manually curated content and static configurations. In practice, constructing and adapting TPRA questionnaires requires domain expertise to align questions with a specific supplier context, making the process increasingly time-consuming as repositories grow. Moreover, repositories accumulate redundancy, inconsistent phrasing, and limited reuse of existing knowledge, further complicating adaptation to organization-specific requirements and evolving risk contexts.


\subsection{Similarity-Based Methods}

Similarity-based retrieval ranks questions or controls by textual relevance. Traditional methods depend on vocabulary overlap and are sensitive to phrasing variation~\cite{robertson2009bm25}. Neural retrieval methods embed queries and documents into dense vector spaces, enabling semantic similarity matching~\cite{reimers2019sbert,karpukhin2020dpr,khattab2020colbert}.

Although effective for finding related questions, these methods do not explicitly represent assessment intent and scope. In practice, this often yields results that are topically related but misaligned with the intended assessment objective, limiting their usefulness for constructing structured TPRA questionnaires. This limitation is also reflected in retrieval-augmented generation (RAG) systems, where retrieval quality strongly affects downstream performance, and semantically relevant but insufficiently structured context may remain misaligned with the underlying task intent \cite{DBLP:conf/nips/LewisPPPKGKLYR020}. Overall, similarity-based retrieval alone is insufficient for capturing the structured semantics required in TPRA.

\subsection{Text Labeling Methods}

An alternative to direct retrieval is to enrich repositories with semantic labels. Unsupervised approaches, such as topic modeling and embedding-based clustering, attempt to discover latent structure without annotation. Classical topic models struggle with short compliance texts~\cite{blei2003lda}, while embedding-based clustering typically assigns each item to a single cluster, limiting the representation of overlapping compliance concepts~\cite{grootendorst2022bertopic}.

Weak supervision generates labels using heuristics or labeling functions~\cite{DBLP:journals/pvldb/RatnerBHBFG17}, but requires a predefined label space and substantial upfront design effort. More recently, Large Language Models (LLMs) have been used to generate human-readable labels without fixed taxonomies~\cite{DBLP:conf/emnlp/TanLLWZ24}. However, LLM-based labeling is costly, sensitive to prompt design, and variable across runs~\cite{DBLP:conf/www/HuangKA23a}.

Overall, existing labeling approaches either lack expressiveness or incur high computational cost, while offering limited support for scalable label propagation across large repositories.

\section{SSSL: Hybrid Semi-Supervised Semantic Labeling Framework}
\label{sec:approach}

This section presents the proposed Semi-Supervised Semantic Labeling framework (SSSL) for TPRA question retrieval. It begins with an overview of the framework, followed by a description of its main components, and concludes with an example demonstrating SSSL in action.

\subsection{SSSL framework overview}
\begin{figure}[htbp]
\centering
\includegraphics[width=0.9\linewidth]{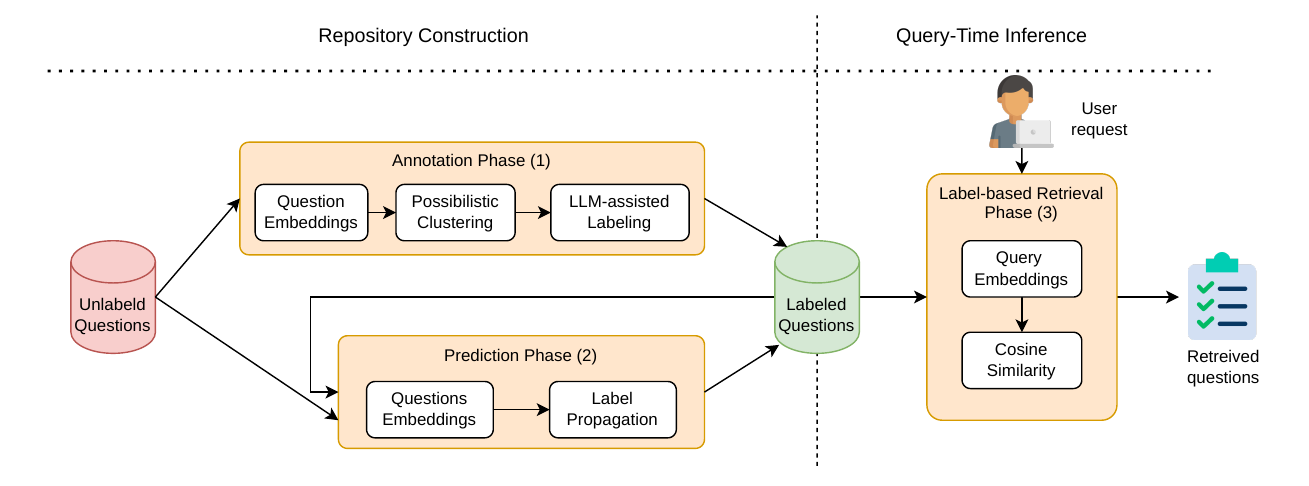}
\caption{Overview of the proposed SSSL framework.}
\label{fig:approach-overview}
\end{figure}

In our effort to investigate retrieval based on a semantic label space as compared to semantic-based retrieval, we propose SSSL framework in Figure~\ref{fig:approach-overview}. Our first step is to construct and generate this label space. Therefore, the first component in SSSL is \textbf{repository construction}, which transforms unlabeled questions into labeled questions. The main objective is to use an LLM for labeling in a way that produces consistent and shared labels across different questions, rather than generating labels independently for each one, while also reducing the overall cost. To achieve this, we provide the model with richer context by presenting multiple questions at the same time, which helps generate shared labels across them while still allowing us to obtain set labels for each question. During an \textit{annotation phase}, we begin by grouping the questions that will serve as input to the model. Specifically, we take unlabeled TPRA compliance questions and group them using semantic embeddings with overlapping cluster membership. An LLM then assigns semantic labels at the cluster level, which are aggregated into multi-label annotations for each question. For example, Q1 defined earlier could belong to a cluster labeled with "Access Control" and another with "Policy Definition" while Q2 could be tied to the same "Access Control" cluster and a third one labeled with "Review Periodicity".

Then, to further reduce cost, we consider label propagation based on the labeled data. The resulting labeled repository enables kNN-based label prediction for new questions without requiring repeated LLM inference during the \textit{prediction phase}.

The second component is \textbf{query time inference}, which supports retrieval in the label space aligned with the supplier context in the user request, enabling the retrieval of relevant TPRA questions during the \textit{label-based retrieval phase}.

\subsection{Annotation Phase}

\label{subsec:annotation}

Let $Q=\{q_i\}_{i=1}^n$ be cybersecurity compliance questions. 
Our objective is to assign each question a multi-label set $L_i \subseteq L$, where $L$ is a free-form label inventory discovered by an LLM.

\paragraph{Question Embeddings.} To group questions into meaningful sets, we first transform each question into a vector representation and then use these vectors to perform the grouping. Each question is encoded as a dense vector $\mathbf{e}_i = f(q_i) \in \mathbb{R}^d$ using a fixed pretrained embedding model. These embeddings enable similarity-based grouping while remaining robust to lexical variation in compliance phrasing.

\paragraph{Possibilistic Clustering.} To form semantically coherent groups, we require effective grouping methods, among which clustering is a natural choice \cite{rokach2005clustering}. However, hard clustering assigns each question to a single group, which is not suitable for our setting, where questions may exhibit multiple semantic aspects. To address this, we adopt Possibilistic C Means (PCM) \cite{DBLP:journals/tfs/BarniCM96,DBLP:journals/tfs/KrishnapuramK93}, which assigns each question an independent membership score for each cluster, thereby allowing flexible and overlapping group assignments. PCM returns a possibility membership matrix
$P_{c,i} \in [0,1]$
where $P_{c,i}$ indicates the degree to which question $q_i$ belongs to cluster $c$.

To construct discrete groups, memberships are thresholded per cluster using elbow (knee) \cite{DBLP:conf/icdcsw/SatopaaAIR11} detection on sorted scores, with fallback to the minimum membership when no elbow is detected. This avoids dataset-specific manual tuning and adapts to the skewed similarity distributions observed in compliance question repositories. Algorithm~\ref{alg:elbow_grouping} summarizes this grouping procedure and produces overlapping clusters $C$ for LLM annotation.

\begin{algorithm}[h]
\caption{Elbow-thresholded PCM grouping}
\label{alg:elbow_grouping}
\begin{algorithmic}[1]
\REQUIRE Questions $Q=\{q_j\}_{j=1}^n$, PCM memberships $P\in \mathbb{R}^{m \times n}$
\ENSURE Overlapping clusters $C$
\STATE $C\leftarrow \emptyset$
\FOR{$c=1$ to $m$}
  \STATE $\mathbf{v}\leftarrow \mathrm{sort}({P}_{c,:},\downarrow)$
  \STATE $t\leftarrow \mathrm{Knee}(\mathbf{v})$
  \STATE $\tau \leftarrow (t=\varnothing)\ ?\ \min(\mathbf{v}) : \mathbf{v}[t]$
  \STATE $S \leftarrow \{q_j \mid {P}_{c,j}\ge \tau\}$
  \STATE ${C}\leftarrow {C}\cup S$
\ENDFOR
\STATE \textbf{return} ${C}$
\end{algorithmic}
\end{algorithm}

\begin{figure}[htbp]
\centering
\includegraphics[width=0.6\linewidth]{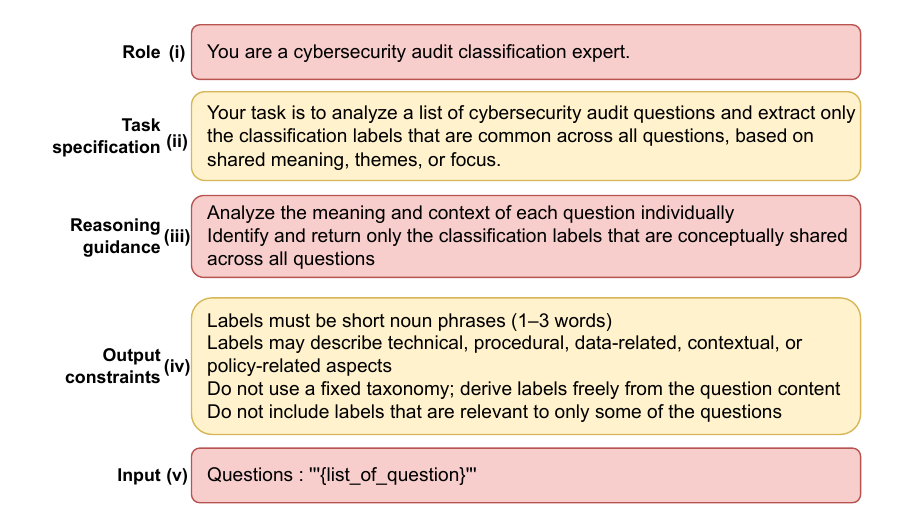}
\caption{Structured prompt template used for cluster-level label extraction.}
\label{fig:prompt}
\end{figure}

\paragraph{LLM-assisted Labeling.}
Rather than labeling individual questions, the LLM is invoked once per cluster to extract shared semantic labels. 
To improve reproducibility and reduce output variance, we use the structured prompt template in Figure~\ref{fig:prompt} \cite{DBLP:conf/emnlp/TanLLWZ24,DBLP:journals/patterns/ChenZLZ25}. The template: (i) assigns a domain role (cybersecurity audit classification), (ii) constrains the task to \emph{cluster-level} labels rather than question-specific details, (iii) enforces comparison across questions to identify shared themes, and (iv) restricts outputs to short noun-phrase labels with no explanations or predefined taxonomies. For each cluster $c$, the LLM returns a label set $L_i$.

Each question inherits the union of labels from clusters containing it:
\[
{L}[q] = \bigcup_{i:\, q \in {C}_i} L_i.
\]

\subsection{Prediction Phase}
\label{subsec:prediction}

To label a new question $q'$ without LLM inference, we apply \emph{k}-nearest neighbors~\cite{DBLP:conf/nips/ZhouBLWS03} in the embedding space. In other words, we apply embedding methods to the labeled questions (or reuse the embeddings computed in the first phase), emphasizing their labels, to obtain embeddings for all questions, and we also compute embeddings for the new question(s). After retrieving the top-$k$ labeled neighbors via cosine similarity, each neighbor votes for its labels, while labels that appear only once are suppressed. 
If no label receives at least two votes, the question is treated as out-of-distribution and deferred to the LLM; otherwise, labels with maximal votes are returned. 
Algorithm~\ref{alg:knn_label_selection} details this procedure.

\begin{algorithm}[h]
\caption{\emph{k}-NN label propagation with OOD fallback}
\label{alg:knn_label_selection}
\begin{algorithmic}[1]
\REQUIRE Labeled questions $Q$, labels ${L}[q]$, encoder $f(\cdot)$, number of neighbors hyperparameter $k$
\REQUIRE New question $q'$
\ENSURE Predicted labels $L(q')$
\STATE $\mathbf{e}'\leftarrow f(q')$
\STATE ${N}\leftarrow \textsc{TopK}\big(\{\cos(\mathbf{e}',f(q))\}_{q\in Q},k\big)$
\STATE initialize counts $S[\ell]\leftarrow 0$ for all labels $\ell$
\FORALL{$q \in {N}$}
  \FORALL{$\ell \in {L}[q]$}
    \STATE $S[\ell]\leftarrow S[\ell]+1$
  \ENDFOR
\ENDFOR
\STATE $s_{\max}\leftarrow \max_{\ell} S[\ell]$
\IF{$s_{\max}<2$}
  \STATE \textbf{return} $\textsc{LLMLabel}(q')$
\ENDIF
\STATE \textbf{return} $\{\ell \mid S[\ell]=s_{\max}\}$
\end{algorithmic}
\end{algorithm}

\subsection{Label-based Retrieval Phase}
\label{subsec:retrieval}

With semantic labels assigned, retrieval operates in label embedding space rather than raw question text, improving alignment with control intent and assessment scope. 
Let ${R}=\{(q_i,{L}[q_i])\}$ be the labeled repository. 
Each label $\ell$ is embedded once as $\mathbf{e}_\ell=f(\ell)$, and a user query $u$ is embedded as $\mathbf{e}_u$, where this query can represent the supplier profile, or the set of questions to be extracted. 
Question relevance is computed by aggregating cosine similarity between $\mathbf{e}_u$ and embeddings of associated labels:
\[
\mathrm{score}(u,q_i)=
\operatorname{Agg}_{\ell\in{L}[q_i]}
\cos(\mathbf{e}_u,\mathbf{e}_\ell),
\]
where mean aggregation is used in practice (max is a viable alternative). 
Top-$r$ questions are returned.

\subsection{SSSL in Action}

We illustrate the proposed pipeline using a small set of assessment questions extracted from \textit{ISO/IEC 27002}. Let $Q=\{q_1,\dots,q_5\}$ denote five unlabeled questions. Each question $q_i$ is embedded as $e_i = f(q_i)$, yielding $E=\{e_1,\dots,e_5\}$ with $e_i \in \mathbb{R}^d$. Applying possibilistic clustering produces the membership matrix:
\[
P =
\begin{bmatrix}
0.82 & 0.79 & 0.21 & 0.08 & 0.25 \\
0.76 & 0.18 & 0.81 & 0.10 & 0.22 \\
0.12 & 0.09 & 0.15 & 0.88 & 0.83 \\
0.71 & 0.14 & 0.19 & 0.22 & 0.78
\end{bmatrix}.
\]

Using elbow-based thresholds $\tau=\{0.79,0.76,0.83,0.71\}$, we obtain overlapping clusters:
\[
C_1=\{q_1,q_2\},\quad
C_2=\{q_1,q_3\},\quad
C_3=\{q_4,q_5\},\quad
C_4=\{q_1,q_5\}.
\]

Cluster-level labeling yields:
$L_1$ = \{\textit{Authentication controls}, \textit{Access governance}\}, 
$L_2$ = \{\textit{Security monitoring}, \textit{Operational oversight}\}, 
$L_3 $ = \{\textit{Operational resilience}, \textit{Service continuity}\}, 
$L_4$ = \{\textit{Incident management}, \textit{Organizational response readiness}\}.
 Each question inherits labels from its clusters. For example: $L(q_1)=L_1 \cup L_2 \cup L_4$.

For a new question $q'$ (from ISO/IEC 27017), we compute $e'=f(q')$ and retrieve its $k=2$ nearest neighbors, assumed to be $q_3$ and $q_1$. Their labels are: $L(q_3)$=\{\textit{Security monitoring}, \textit{Operational oversight}\}, 
$L(q_1)$=\{\textit{Authentication controls}, \textit{Access governance}, \textit{Security monitoring}, \textit{Operational oversight}, \textit{Incident management}, \textit{Organizational response readiness}\}. 
 Labels are predicted by majority voting. The most frequent labels are: $L(q')$=\{\textit{Security monitoring}, \textit{Operational oversight}\}.

\section{Evaluation}
\label{sec:evaluation}

We evaluate whether SSSL reduces reliance on repeated LLM-based labeling for compliance self-assessment questionnaires while enabling retrieval via reusable semantic labels rather than question-level similarity. To support reproducibility, we release an open-source implementation for research and non-commercial use, datasets, and evaluation scripts at: \url{https://github.com/NourEldin-Ali/semi-supervised-semantic-labeling}.

\subsection{Experimental Setup}

\textbf{Datasets.}
We evaluate SSSL on cybersecurity compliance self-assessment questionnaires, a common artifact in governance, risk, and compliance workflows. Publicly available TPRA-style questionnaire datasets are scarce and we include only the CSA CAIQ benchmark as a real-world dataset.\footnote{\url{https://d1.awsstatic.com/whitepapers/compliance/CSA_Consensus_Assessments_Initiative_Questionnaire.pdf}}

To extend the evaluation across multiple standards, we construct synthetic questionnaire datasets using ChatGPT,\footnote{\url{https://chatgpt.com/}} guided by public control scope descriptions from widely used standards, including ISO/IEC~27001, ISO/IEC~27002, ISO/IEC~27017, ISO/IEC~27018, and ISO/IEC~27701. These datasets are designed to resemble TPRA-style repositories by expressing controls as short natural-language assessment questions. They preserve two characteristics commonly observed in real questionnaire repositories: strong semantic overlap and the absence of a fixed, predefined labeling taxonomy. Table~\ref{tab:datasets} summarizes the datasets used in our evaluation.

\begin{table}[htbp]
\centering
\small
\begin{tabular}{lccc}
\hline
\textbf{Dataset} & \textbf{\# Questions} & \textbf{\# Control Domains} & \textbf{Type} \\ \hline
CAIQ & 260 & 17 & Real-world \\
ISO/IEC 27001 & 200 & 22 & Synthetic \\
ISO/IEC 27002 & 250 & 59 & Synthetic \\
ISO/IEC 27017 & 200 & 39 & Synthetic \\
ISO/IEC 27018 & 200 & 40 & Synthetic \\
ISO/IEC 27701 & 200 & 40 & Synthetic \\ \hline
\end{tabular}
\caption{Summary of datasets used in the evaluation.}
\label{tab:datasets}
\end{table}

\textbf{Baselines, Tasks, and Metrics.}
To align our evaluation with real-world scenarios, we conduct experiments through two phases. In Phase I, we simulate a fresh knowledge base construction where the TPRA-style repository contains questions with no associated semantic labels. We want to compare in this phase the efficiency and effictiveness of per-question LLM-annotation (\emph{baseline}) versus the first step of SSSL (\emph{SSSL-I}), in which semantic labels are generated for clusters and by extension for each question inside. In Phase-II, another questionnaire dataset (\emph{extension dataset)} is added to the repository and we compare the efficiency and effectiveness of another round of per-question LLM annotation (\emph{baseline}) versus the second step of SSSL (\emph{SSSL-II}), in which labels are predicted using kNN propagation based on embedding similarity. It is to be noted that during Phase-II, we will attempt to apply transfer learning from one domain (e.g., ISO/IEC 27002) to another domain (e.g., ISO/IEC 27017), which partially overlaps with the former, in order to assess the transferability of domain labels across domains.



We evaluate two tasks: (1) label quality and (2) question retrieval. Label quality is assessed through three metrics: \emph{correctness}, \emph{generalization}, and \emph{consistency}. \emph{Correctness} assesses a label's semantic accuracy and terminological relevance for cybersecurity compliance questions. \emph{Generalization} assesses whether a label captures the underlying control intent and can be applied across questions with similar meanings, rather than being limited to a single wording instance. \emph{Consistency} assesses whether labels are assigned consistently, uniformly, and comparably across semantically related question groups. Retrieval determines whether semantic labels improve the identification of questions that have the same control scope. Additionally, we report execution time, token usage, and energy consumption\footnote{Energy measured using CodeCarbon: \url{https://codecarbon.io/}.}.

We conduct both automatic and human evaluations. Automatic evaluation relies on LLM-as-a-Judge \cite{DBLP:journals/corr/abs-2411-15594}, which compares SSSL and the baseline on a 1--5 scale for each metric. In this evaluation, all datasets were used for Phase-I except ISO/IEC~27002 which was kept as the extension dataset for Phase-II. Human evaluation involves 25 participants (13 master’s students, 7 PhD students, and 5 engineers). Phase-I was executed on ISO/IEC~27002 and ISO/IEC~27017 served as the extension dataset.

\textbf{Implementation.} 
We use \texttt{GPT-5.1} for all LLM-based components and \texttt{text-embedding-3-large} for question embeddings. The same models are used across all stages of the pipeline and kept fixed throughout the evaluation to ensure consistency and comparability.

\subsection{Results \& Discussion}
We observe consistent relative trends across all datasets and settings. Average is reported across datasets.\footnote{Due to space limitations, we report only the average results in this paper; the full results are available in our GitHub repository.}

\paragraph{Semantic Label Quality.}
The results of the automatic evaluation for the two phases and across the three metrics are given in Table~\ref{tab:label_quality}. For each metric, we report the the average results across the datasets used in the phase. The table shows that LLM-only labeling or baseline achieves the strongest correctness (4.8) and consistency (4.8). The cluster-level LLM phase of SSSL (SSSL-I) that leverages shared context within semantically coherent groups, produces comparably consistent labels (4.7) with comparable generalization (4.3) under reduced cost. Finally, the kNN-based propagation phase (SSSL-II) maintains identical consistency (4.7) while reducing correctness (1.8) and generalization (3.4), indicating that propagation is efficient but can be sensitive to boundary cases where nearest-neighbor transfer is less precise.

\begin{table}[htbp]
\centering
\begin{tabular}{cccc}
\hline
\textbf{Method}           & Correctness & Generalization & Consistency \\ \hline
baseline                       & 4.8         & 4.3            & 4.8         \\
SSSL-I         & 3.5         & 4.3            & 4.7         \\
SSSL-II          & 1.8         & 3.4            & 4.7         \\ \hline
\end{tabular}
\caption{Semantic label quality comparison evaluated using an LLM-based judge. Higher is better.}
\label{tab:label_quality}
\end{table}

In the human evaluation, where participants assigned scores from 1 to 5, labels generated with SSSL-I achieved a consistency score of 4.01, correctness of 3.68, and generalization of 4.19. In comparison, LLM-generated labels in Phase-I obtained higher scores in consistency (4.148) and correctness (4.368), but slightly lower generalizability (3.944). SSSL-II, where question labels are propagated from ISO/IEC 27002, provided scores of 3.832 in consistency, 3.744 in correctness, and 4.100 in generalizability. By contrast, LLM-generated labels during Phase-II achieved higher consistency (4.296) and correctness (4.528), with lower generalizability (3.928).

\paragraph{Computational Efficiency and Scalability.}
We next compare computational cost between SSSL and baseline during the automatic evaluation. Table~\ref{tab:cost} summarizes the number of labels produced, energy usage, token consumption, and runtime for the two phases.

\begin{table}[htbp]
\centering
\small
\setlength{\tabcolsep}{2pt}

\renewcommand{\arraystretch}{1.05}
\begin{tabular}{cccccc}
\textbf{Phase}            & \textbf{Method} & \textbf{Labels \#} & \textbf{Energy (kwh)} & \textbf{Tokens} & \textbf{Time (s)} \\ \hline
\multirow{2}{*}{Phase-I}  & Baseline        & 621                & 0.003                 & 55848           & 314               \\
                          & SSSL-I          & 106                & 0.002                 & 29471           & 197               \\ \hline
\multirow{2}{*}{Phase-II} & Baseline        & 670                & 0.003                 & 63640           & 368               \\
                          & SSSL-II         & 18                 & 0.000002              & 0               & 0.22 \\ \hline          
\end{tabular}
\caption{Computational cost comparison. Lower is better except for Labels \#.}
\label{tab:cost}
\end{table}

The baseline incurs substantial overhead due to repeated per-question inference (3.5x slower overall). During Phase-I, SSSL-I reduces LLM usage by concentrating annotation on a smaller set of representative clusters (106 labels), cutting token usage from 55{,}848 to 29{,}471 (a 52.9\% reduction) and runtime from 314\,s to 197\,s (a 62.7\% reduction). Crucially, SSSL-II eliminates LLM calls during Phase-II (0 tokens) and runs in 0.22\,s, yielding a \(\sim\)1673\(\times\) speedup over baseline annotation (368\,s \(\rightarrow\) 0.22\,s) and a \(\sim\)1500\(\times\) reduction in energy (0.003\,kWh \(\rightarrow\) 0.000002\,kWh). These results highlight that SSSL can support large-scale deployment by separating expensive label discovery from low-cost label assignment.

\paragraph{Question Retrieval Quality.}
We next evaluate downstream \emph{question retrieval}: given a high-level assessment intent, the system retrieves and ranks candidate questions that best match the intent.  
Performance is scored from 0 to 100 using LLM-as-Judge, where 100 denotes a perfect match.  
We consider three queries: \textbf{Q1} (\emph{Backup Frequency \& Retention}), \textbf{Q2} (\emph{Monitoring \& Incident Response}), and the composite \textbf{Q3} combining both domains.

\begin{table}[htbp]
\centering
\small
\begin{tabular}{cccc|c}
\hline
\textbf{Method}              & \textbf{Q1} & \textbf{Q2} & \textbf{Q3} & \textbf{Overall} \\ \hline
BM25                         & 28          & 78          & 70          & 58               \\
Semantic similarity          & 58          & 90          & 62          & 70               \\
Labeling-based Similarity & 65          & 88          & 72          & 75               \\ \hline
\end{tabular}
\caption{Question retrieval quality evaluated using an LLM-based judge. Higher is better.}
\label{tab:retrieval_quality}
\end{table}

Table~\ref{tab:retrieval_quality} shows that BM25 performs unevenly, with strong results on \textbf{Q2} (78) but weak performance on \textbf{Q1} (28) due to lexical variation.  
Embedding-based similarity improves the overall score over BM25 (70) yet drops on the composite \textbf{Q3} (62), indicating partial intent matching.  
Incorporating semantic labels yields the best overall performance (75) and improves both \textbf{Q1} and \textbf{Q3} over the two other methods. On \textbf{Q2}, it improves over BPM25.

For human evaluation, participants assessed only \textbf{Q1}, the lowest-scoring query in the automated evaluation.  
Semantic similarity achieved an average score of 3.24, while label-based retrieval reached 3.95.

\paragraph{Discussion.}
Taken together, these findings show that separating semantic label discovery from large-scale label assignment provides an effective and scalable alternative to direct LLM-only labeling.
With respect to \textbf{RQ2}, the proposed SSSL pipeline demonstrates that semantic labels can be induced efficiently through a cluster-level LLM stage and then propagated at scale through kNN-based assignment.
In contrast to per-question LLM annotation, this decoupled design substantially lowers operational cost by restricting expensive LLM inference to representative clusters during Phase-I, reducing token consumption by roughly 50\% and runtime by about one third, while completely removing the need for LLM calls at deployment time.

Human evaluation reinforces these conclusions.
Across ISO/IEC~27002 and ISO/IEC~27017, LLM-generated labels achieve slightly higher correctness and consistency, whereas SSSL preserves strong overall label quality and higher generalization, particularly in ISO/IEC~27002, indicating that cluster-level abstraction better captures control intent.
When transferring labels across standards (ISO/IEC~27002 $\rightarrow$ ISO/IEC~27017), SSSL maintains competitive generalizability with only moderate reductions in correctness and consistency, suggesting practical viability under partial taxonomy overlap.
These findings further support \textbf{RQ2} by showing that reduced dependence on repeated LLM inference remains feasible without fully sacrificing semantic usefulness.

In relation to \textbf{RQ1}, human assessment of retrieval further supports the benefit of the proposed semantic labeling strategy.
For the most challenging query (\textbf{Q1}), semantic similarity scores 3.24, while label-based retrieval reaches 3.95, confirming that reusable semantic labels provide a clearer abstraction for intent matching (Table~\ref{tab:retrieval_quality}).
This suggests that semantic labels improve retrieval alignment beyond raw semantic similarity alone, especially when matching depends on shared control intent rather than surface-level semantic relatedness.

The lower correctness in SSSL-II (Table~\ref{tab:label_quality}; 1.8 vs.\ 3.5 in SSSL-I) reveals a scalability limitation of SSSL across partially overlapping domains; when overlap is insufficient, label transfer becomes ineffective, resulting in semantic drift and reduced transferable label diversity. This effect reflects cross-standard mismatch and label-space contraction (Table~\ref{tab:cost}; 18 vs.\ 106 labels).
At the same time, these results also address \textbf{RQ3} by clarifying the trade-off between scalability and semantic precision.

Despite this limitation, kNN remains practically valuable by enabling near-instant labeling (0.22\,s; 0 tokens) while preserving useful generalization and retrieval performance, supporting rapid deployment and iterative questionnaire updates.
Overall, the findings answer \textbf{RQ3} by showing that the proposed framework offers a favorable balance between cost, scalability, and downstream utility, even if performance is not preserved equally across all quality dimensions.

\subsection{Threats to Validity}
Our evaluation focuses on cybersecurity compliance questionnaires used in TPRA workflows. 
Although representative of real industrial practice, results may differ in other governance or assessment domains with distinct terminology, granularity, or structure. 
Evaluating SSSL across additional industrial settings is therefore left for future work.

We also rely on a single LLM for both semantic label discovery and LLM-based evaluation, and on a single embedding model for question representation. As a result, the observed performance may depend in part on the specific language and embedding models used, including the embedding dimensionality, which can affect clustering quality, nearest-neighbor propagation, and retrieval behavior. Future work should therefore include both multi-LLM evaluation and experiments with different embedding models and embedding dimensions to better assess robustness.

Finally, kNN propagation performance depends on overlap between source and target questionnaires.
Our experiments use a cross-standard transfer setting with partial taxonomy alignment, which can reduce correctness and label diversity.
Evaluating kNN under the same-distribution and incremental real-world scenarios would help disentangle cross-standard mismatch from intrinsic kNN limitations. Future work may mitigate cross-standard degradation through label-level grouping that expands transferable semantic structure without increasing initial LLM annotation cost.

\section{Conclusion}
\label{sec:conclusion}
We presented a hybrid framework for semantic labeling in TPRA compliance cybersecurity question repositories that combines unsupervised semantic grouping with selective use of LLMs. By inducing interpretable labels and propagating them efficiently, the approach reduces reliance on repeated LLM inference while maintaining labeling quality. Beyond annotation efficiency, the induced labels enable more focused and adaptive TPRA questionnaire generation by making control domains and assessment intent explicit, improving question selection and reducing redundancy. Future work will explore automated question answering using organizational knowledge and semantic labels.

\section*{Acknowledgments}
This work was supported by Bpifrance\footnote{\url{https://www.bpifrance.fr/}} within the framework of the French national public-private partnership project EVA2026.

\bibliographystyle{splncs04}
\bibliography{biblio}

\end{document}